\documentclass{elsart}

\usepackage{epsfig}
\usepackage{url}

\begin{document}

\begin{frontmatter}
\title{Implementing an Agent Trade Server}


\author{Magnus Boman}
\address{Swedish Institute of Computer Science (SICS) AB and The Royal Institute of Technology, Kista, Sweden}
\author{Anna Sandin}
\address{Swedish Institute of Computer Science (SICS) AB, Kista, Sweden}

\begin{abstract}
An experimental server for stock trading autonomous agents is presented and made available, together with an agent shell for swift development. The server, written in Java, was implemented as proof-of-concept for an agent trade server for a real financial exchange. 
\end{abstract}

\end{frontmatter}


            

            
            
            
            
            

\section{Introduction}
\begin{figure}
\centering
\includegraphics[width=14cm]{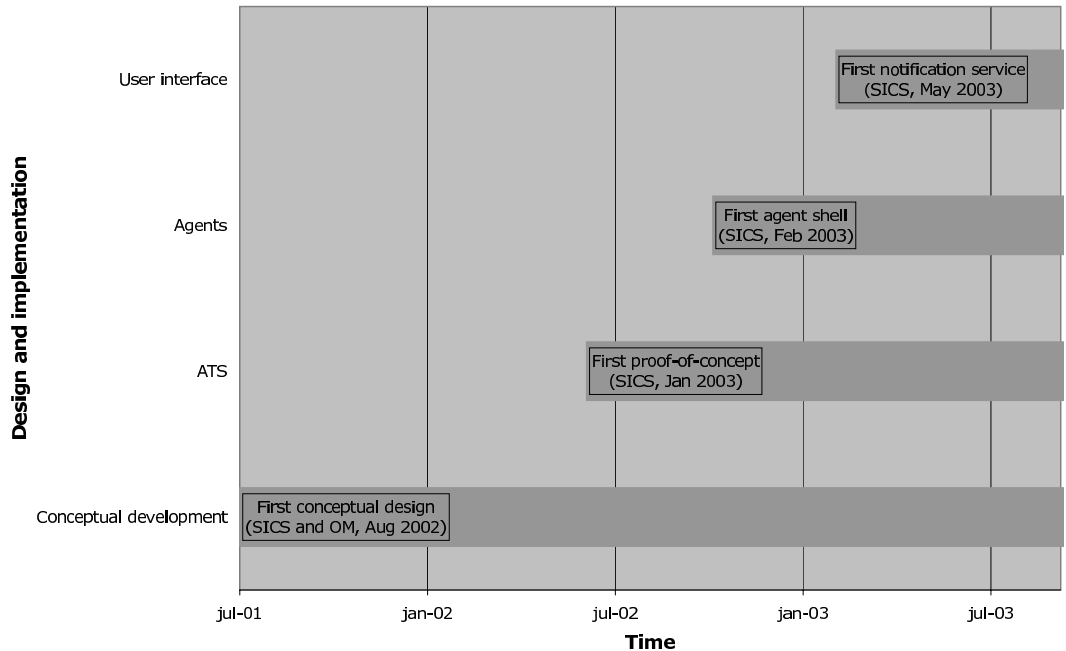}
\caption{The four phases of the ATS implementation project, with the most fundamental at the bottom, and with all four still running. The Gantt schema also indicates key breakthroughs in each of the four phases. The second conceptual design will hopefully be made at OM, or at some other ITC provider for F/X software, while in itself an interesting academic exercise. The first ATS was made publicly available in January 2003, but the permission to run agents on the ATS was restricted to select agent developers at SICS and at two Swedish universities. These developers came up with a first shell in January, which was refined over the following months, and sophisticated agents are now under development. As a result of a small empirical study, not reported on here, the first user service---a trade notification service for smart phones---was recently implemented.}
\label{gantt}
\end{figure}

There are always investors seeking to place orders more complicated than can be accepted by the software of a financial exchange (F/X) system, regardless of its current level of sophistication. To some of these investors, the agent metaphor is a means to implementing combinatorial, temporal, contingent, or otherwise complex orders. In general, a trading agent is a piece of encapsulated software that codes the preferences of its owner. In theoretical research, trading agents have been used in idealized games~\cite{RuMiPa94}, artificial markets~\cite{PaArHoLeTa94}, and competitions such as the Trading Agent Competition~\cite{Gr03} (see \url{http://www.sics.se/tac/}). To economists, trading agents are speculating noise traders (traders with non-rational expectations and potentially zero intelligence)~\cite{DeShSuWa91}, often subjected to the Efficient Market Hypothesis and the Rational Expectations Hypothesis~\cite{LuAu02}, since the empirical evidence against the accompanying assumptions are directed almost exclusively towards human traders~\cite{Lu95}. In the system of trading agents we consider, the agents interact with the same order book as the human traders (cf. \cite{BoJoLy01}), resulting in a system only slightly more complex than an F/X is today, but a system that is extremely difficult to analyze (cf., e.g., \cite{JeHaHuJo01}). From an academic perspective, the introduction of agents in the actual exchange of real stocks is a formidable challenge in that it prompts research in real-time control, agent programming, decision analysis, user interaction, privacy, and security. Its theoretical models must go beyond agent-based computational finance~\cite{Le00} and agent-based computational economics~\cite{Te02}, and already simulation studies are highly complex~\cite{Le97}~\cite{LuMa99}~\cite{BeBo01}~\cite{DoJo02}. From a commercial perspective, the concept is promising in that it enables implementation of new services in the service portfolio of the F/X. A vast range of commercial efforts have helped renew and refine the concept of an electronic F/X, including ambitious liquidity-moving approaches, such as OptiMark~\cite{LuRi97}~\cite{ClWe98}~\cite{GeTeWaWa03}. This fact notwithstanding, there has been no technical platform for real stock trading agent development and no special-purpose server generally available. We rectify this matter by making available our documented Java code~\cite{Sa02} for such a platform. We have also implemented an experimental server for supporting agent trading of stocks, an Agent Trade Server (ATS). The conceptual development~\cite{LyBo03} was initiated in 2001 (cf. Fig.~\ref{gantt}) in cooperation with OM, the world's largest supplier of software to stock exchanges~\cite{Sa01}, and OM has filed a U.S. patent on the ATS concept~\cite{OM02}. The ATS was built not with commercial operation in mind, but as proof-of-concept. That said, OM's trading package SAXESS and its third party APIs provided for much inspiration. We have not designed the ATS for any particular market design. Instead, all agents must abide by the rules of the F/X. The submarket emerging from ATS operations will depend on the agents' business logic, including their decision-analytic~\cite{EkDaBo97} and normative modeling~\cite{Bo99} of the market, which we have previously investigated. We have no illusion of ATS traffic accounting for more than one per cent of any stock exchange before 2010, and whether or not investors will find the ATS attractive will to a large extent depend on pricing and other commercial or political concerns.

The following section briefly describes the role that software, and agent software in particular, plays at an electronic F/X. In Section~\ref{sec:architecture}, the architecture of our ATS is explained. In Section~\ref{sec:shell}, the issue of logging and other forms of keeping history is highlighted. The various business roles, i.e. those interested in the logs, are also discussed. We end with our conclusions and directions for future research. 

\begin{figure}
\centering
\includegraphics[width=12cm]{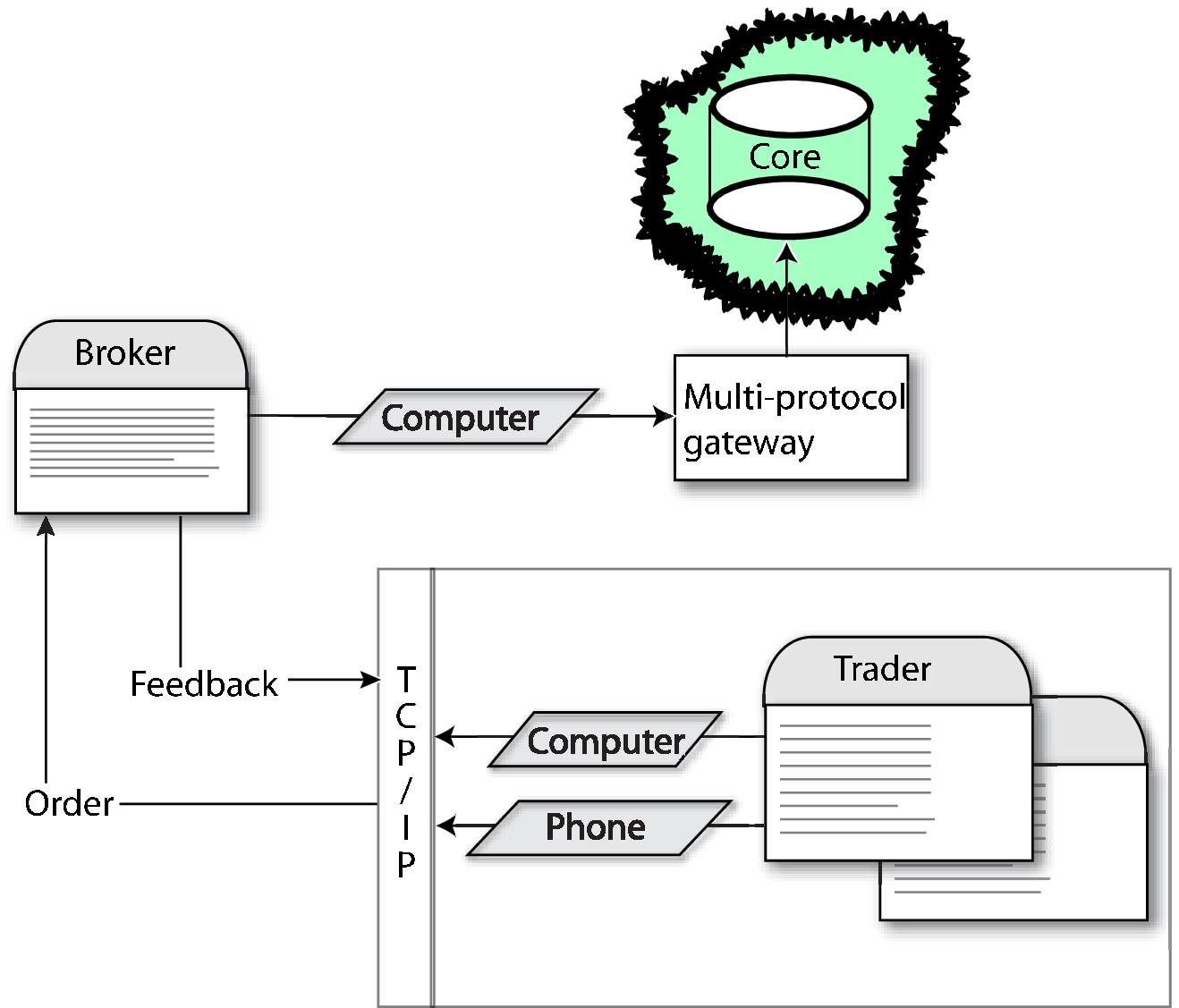}
\caption{A schematic view of the current trade information flow. For simplicity, dissemination procedures are not shown.}
\label{flow}
\end{figure}

\section{Development of Financial Exchange Software}
\label{sec:third}
Trading of stocks is normally done through a broker, as shown in Fig.~\ref{flow}. The trader can be a professional investor, a daytrader, or any other person using stocks as a means to invest. The broker could be a bank, an Internet broker, or some other financial institution. The figure is simplified: orders can also be placed by walking into a bank and filling out a form, IP phones are no requirement, the computer used is often a PDA, etc. Each broker communicates with the core of the F/X via a gateway, over leased lines (at Stockholmsb\"orsen, the Swedish stock exchange, backed up by an ISDN connection, as well as {\it OMnet}, a fibre-optic MetroLAN option). In Fig.~\ref{flow}, the dissemination of information from the F/X to the outside world is not shown. Suffice to say here that stock exchange dissemination is an elaborate system of latencies and routing, in order to secure fairness, and that such information is an important part of the feedback from core to broker, and from broker to trader. 

\begin{figure}
\centering
\includegraphics[width=12cm]{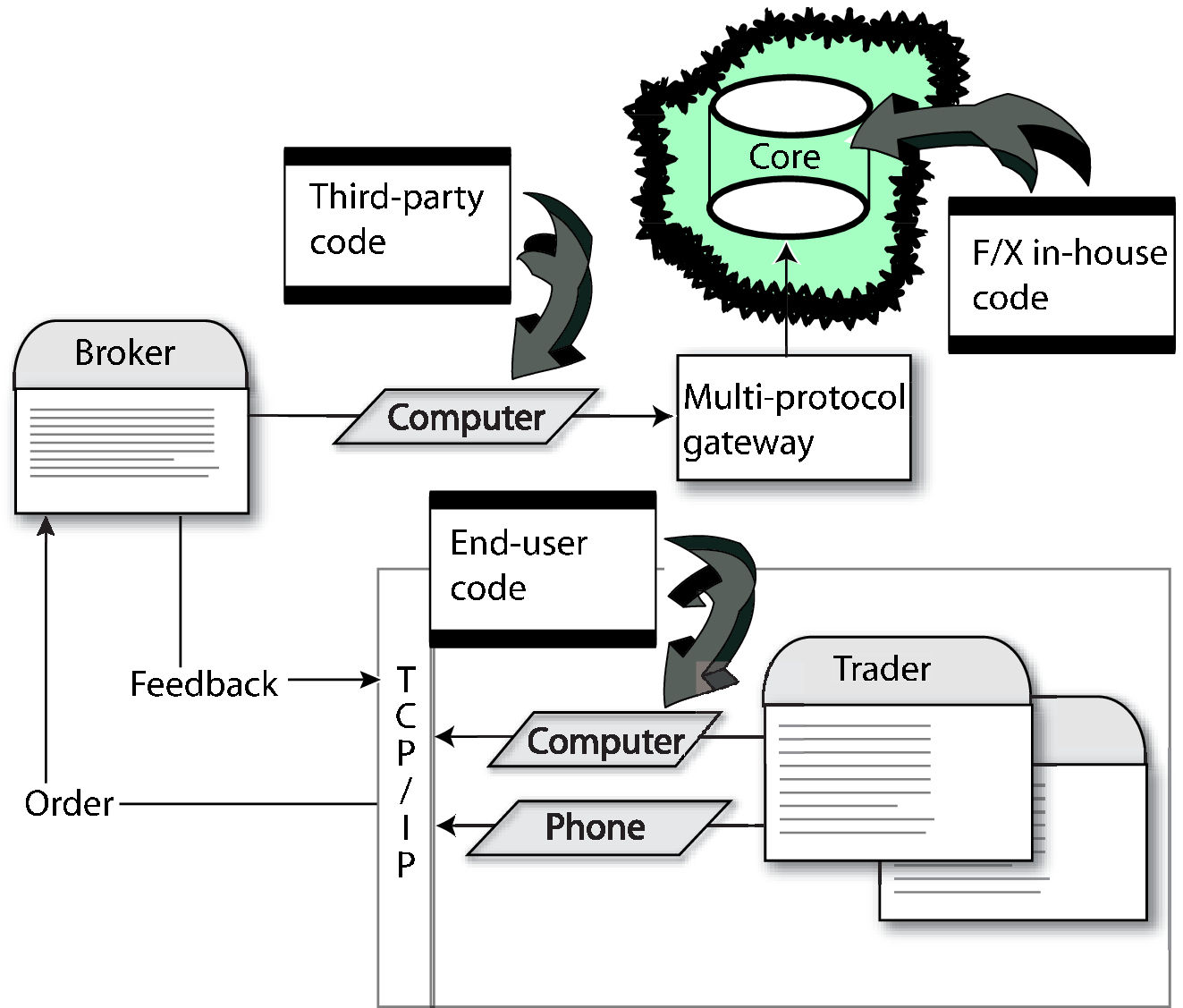}
\caption{Three kinds of code development.}
\label{dev}
\end{figure}

Suppliers of software to an electronic exchange system are basically of two kinds: those that provide software serving the core (the order book), and those that provide software for those interacting with the core (the traders). In most countries, the software supplier in the first category is also competing in the second category. In Sweden, for example, OM provides all of the software for the core, and they also provide the SAXESS and CLICK (for the derivatives market, which we will neglect entirely at this stage) software trading packages, as well as the SECUR software clearing and back office package. The bulk of the software aimed at traders and brokers is however developed by third party software houses. We would like to complicate Fig.~\ref{flow} slightly (see Fig.~\ref{dev}) by considering individuals as service provisioners~\cite{Es03}, thus introducing a third kind of code developer. We anticipate that end-user code development in the form of trading agents will be an increasingly important part of the future F/X service portfolio.

\begin{figure}
\centering
\includegraphics[width=9cm]{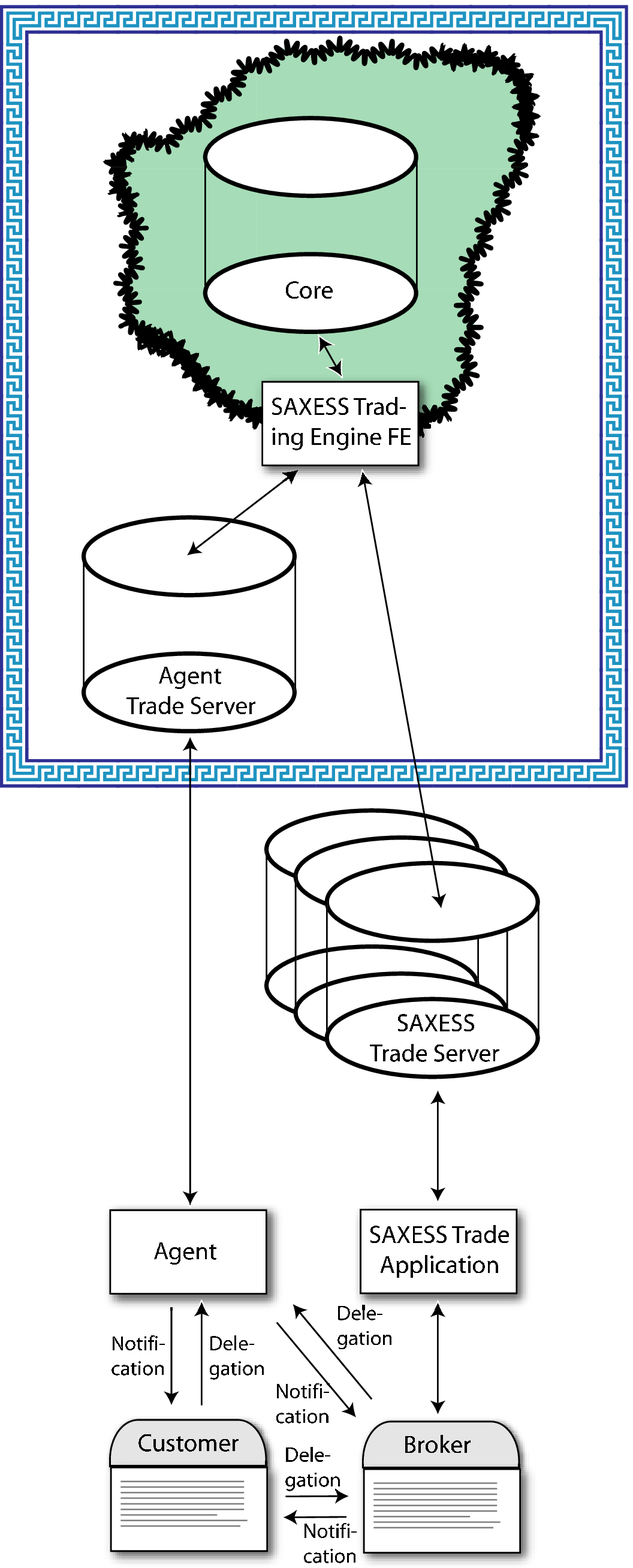}
\caption{The ATS as a complement to ordinary electronic trading. }
\label{sax}
\end{figure}

The third party code pertains to the trading applications used. In order to use the SAXESS system today, the broker needs a member SAXESS trade application. Each such application must be certified, to guarantee smooth operation alongside all other running applications. The OM application offered is {\it SAXESS Trade}, an NT-based client-server solution with the trading data being maintained on an SQL server. There are currently about two dozen certified alternatives. Each application connects to a SAXESS trade server, which is in turn connected to the core. The transaction protocol used is XTP (open eXchange Transaction Protocol) 2.40, an OSI Layer 7 protocol. The Session Layer protocol used is XMP (eXchange Message Protocol). The transactions involving the core are governed by a C program, essentially matching the stack of sell orders with the stack of buy orders. It is important to note that the matching is deterministic. The ATS concept is not a suggested alternative to the SAXESS (or any other) trade server currently in use, but a complement. In Fig.~\ref{sax}, the ordinary electronic trading is represented by SAXESS. Upward arrows denote the placing of orders while downward arrows denote feedback. Each trade application uses zero or more distributed instances of the SAXESS Trade Server, a freeware available to all registered traders. In practice, the ATS will also contain an instance of the SAXESS Trade Server, since the latter contains the latest API for communicating with the core. All communication with the core goes via the Front End (FE). The wall indicates the physical and logical boundary between the market owner and the outside world, while the barbed wire surrounding the core illustrates a further level of protection. The shadow backup system is not shown. 

Note that brokers may still be used, for example for legal reasons, when delegating trading to an agent. While delegating the right to place orders in itself does not introduce non-determinism in the marketplace, there is nothing that {\it per se} stops an agent developer to introduce non-determinism in the business logic of the agent. Security control of agent code is done at the time of agent certification and testing, and also online in the ATS.

In SAXESS, each trade application must also connect to a dedicated Distributed Dissemination Server (DDS). With an ATS running together with a SAXESS trade server, the owner of the F/X has the option to either route all traffic feedback through the existing DDS channels, or to complement the DDS with ATS traffic feedback. In particular, the agents running on the ATS could subscribe to ATS traffic feedback of various sorts, and at variable cost. Even non-professional traders are no strangers to varying levels of sophistication of the individually specified trading constraints built on feedback. The stop loss service offered by Internet brokers (such as Avanza; see \url{http://www.avanza.com}) is a case in point.


The ATS concept provides the third party software houses with a business opportunity, viz. to provide code for the agents trading on the ATS. This involves coding investor preferences and encapsulating these into agents, whereafter the agents are placed on the ATS, without disclosure. Preference elicitation is as always difficult, but suffice to say here that investors will come in many varieties, from those that have developed advanced agents on their own, to those that have no clue on agent programming. A template agent can be coded based on the investor filling out a fairly simple questionnaire. The size and structure of such template agents is simple, and the first agent shell for our ATS has already been written~\cite{JoPo03}. After the agent is placed on the ATS, there remains the challenge of providing software services for notification and (depending on security issues) possibly for termination of agents. Investors no doubt require interfaces for a variety of modalities, and will in the future include roaming~\cite{Boetal02}. 

\section{An Agent Trade Server Architecture}
\label{sec:architecture}

\begin{figure}
\centering
\includegraphics[width=10cm]{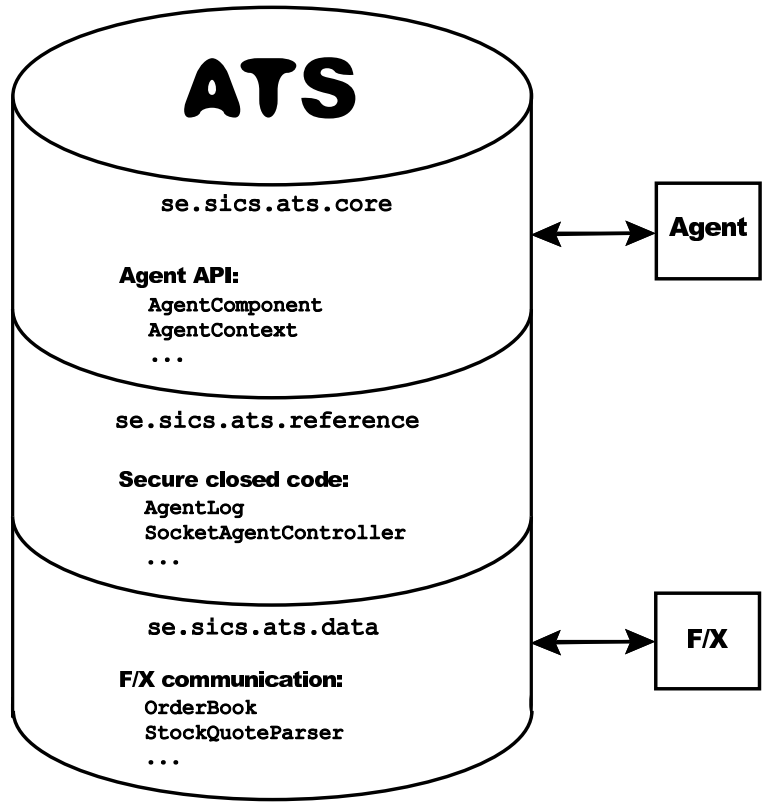}
\caption{The three ATS packages implemented.}
\label{ats}
\end{figure}

We developed the ATS in-house at SICS, in order to allow for free experimentation with agent code and simultaneously adjusting and adding code to the server. A jar-file containing an agent can be run locally or remotely. Each trading agent executes in its own thread. The three server packages implemented (cf. Fig.~\ref{ats}) are described in turn below. The choice of Java was motivated, among several factors, by smooth compatibility with small devices, such as Java-enabled cell phones. While there is no theoretical reason to limit any sandbox environment for agent development to only one language, Java currently seems the most practical first choice. For instance, Java's policy files are useful for defining the boundary between ATS and agent, and \texttt{java.security} allows for basic signing of agent code.

\subsection{se.sics.ats.core}
This package represents the interface towards the agents, and is distributed to agent developers. Initially, the list of agent developers will most likely overlap considerably with the list of certified SAXESS members, and the principles for testing similar. The next step is then to automate the certification process so that individual agent developers can submit agents to the ATS. Since there are legal and practical reasons for the market owner not to certify every single developer, there is room here for traditional as well as agent brokerage. The class \texttt{AgentComponent} specifies all ATS interfacing demands on the agent. For instance, the agent must have functionality for starting and stopping. The class \texttt{AgentContext} is the agent's handle on the server. \texttt{AgentContext} is inspired by and adheres to the SAXESS Trade API for third party code development, based on Microsoft's Component Object Model (COM).

\subsection{se.sics.ats.reference}
This package contains the implementation of the ATS itself. It contains log handlers and socket tranceivers.

\subsection{se.sics.ats.data}
This package represents the interface towards the stock exchange. This is where the ATS will connect to the core of the F/X. For now, it contains a parser which, like the connection, can easily be replaced by another. It now parses data disseminated through the WWW (more specifically, it parses \url{http://www.stockholmsborsen.se}). The ATS thus uses real stock exchange data, albeit with a delay of up to 15 minutes. We have limited its capability at the outset to the most traded stocks on the Stockholmsb\"orsen A-list. The package also simulates an order book. 

\section{Agent Shell Execution}
\label{sec:shell}
We can now specify the ``Agent'' rectangle from Fig.~\ref{ats} above by explaining the agent shell. We first turn to some more abstract aspects of ATS management, however.

\subsection{Roles}
In our proof-of-concept implementation work, we have felt a need for a division of labor between agent developers and ourselves: the ATS developers. In order to leave the core software unchanged---a requirement from OM, and most likely from any market owner---we also need to define the role of ATS administrator. The regulation of trade in SAXESS is today done through surveillance of the Trading Engine (SaxView), as well as through the official trading control and supervisory functions units. These units operate on the core and will be used also for agent trading. In addition, the ATS administrator must store and sign certificates for authenticated agent code developers.

In addition, there are new business roles in the future scenario of commercially available ATS services, perhaps including dedicated agent brokers. We have enabled this new business role by implementing an object relating agent brokers and their agents. This object maintains references, deducts brokerage fees, and supplies data for notification services. For security reasons, the agent broker object is placed in the same execution environment as the ATS. The only agent manipulation granted the object is the killing of an agent thread, should this become necessary.

\begin{figure}
\centering
\includegraphics[width=11cm]{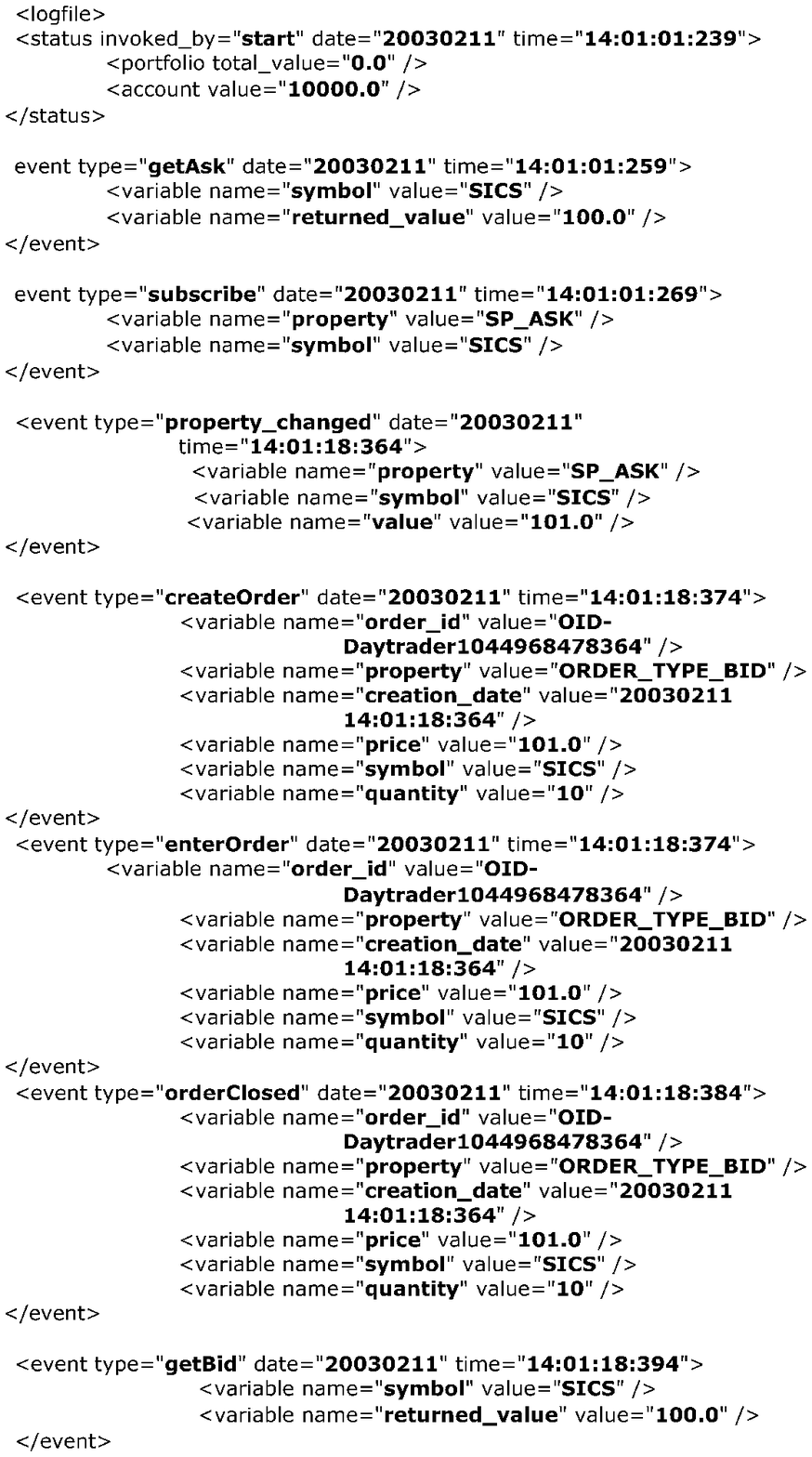}
\caption{Excerpt from a log of Daytrader, a simple agent running on the ATS.}
\label{log}
\end{figure}

\subsection{Logs}
The ATS administrator must monitor the actions of all agents on the ATS, and activity logs are therefore needed. These logs are likely to be processed only automatically, but in case of system failure or suspected illicit agent action, manual inspection may be required. For this reason, the responsibility for logging should not be distributed, but be the responsibility of the ATS administrator. Each agent should be allowed read and write access to its own action logs, since logs provide interesting information on top of the generally available trade statistics. Logged information could be used for machine learning, which could occur offline or online, depending on the agent business logic. An excerpt from a log of a simple agent, to be discussed further in the next subsection, is shown in Fig.~\ref{log}. The full source code of the agent, as well as the full log, has been published~\cite{JoPo03}. Since all stakeholders are potentially interested in parsing the log, it is in XML format. For instance, the ATS administrator might like to find the agent responsible for some illegal action on the ATS. Analogously, an agent broker might seek to attribute errors to the ATS implementation, using logs as evidence. The error log of Java exceptions and \texttt{String} objects is kept separate from all other logged data. 

\subsection{se.sics.ats.agent}
\begin{figure}
\centering
\includegraphics[width=10cm]{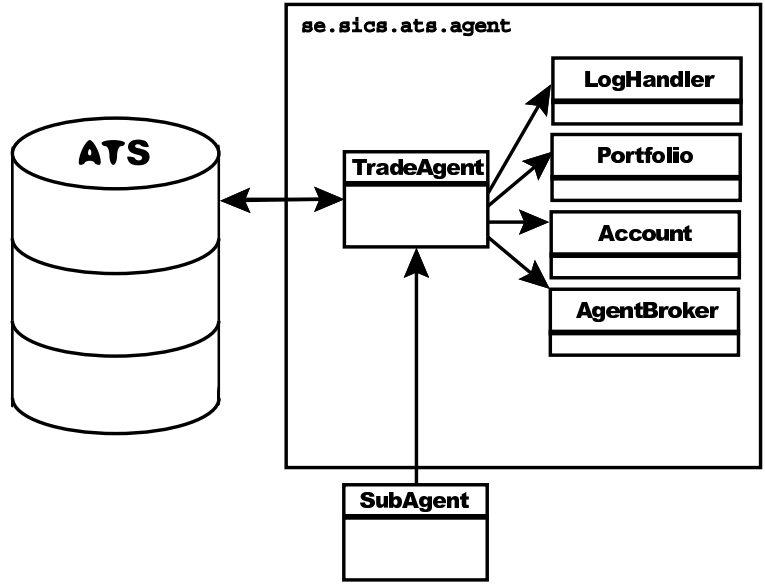}
\caption{The agent shell package structure.}
\label{agent}
\end{figure}

The structurally designed agent shell packages are intended as support for ATS agent development, and their JavaDoc is publicly available~\cite{JoPo03b}. We give an abstract overview here of the low coupling design, so that the role of the shell in our implementation of the ATS is made clear. Agents extend the \texttt{abstract TradeAgent} class (see Fig.~\ref{agent}), which handles all ATS interaction, including the \texttt{final} methods \texttt{initialize} and \texttt{start}, invoked by the ATS~\cite{Sa02}. The agent (\texttt{SubAgent}) gets information about stocks through various \texttt{get} methods, or by subscribing to stock changes (cf. Fig.~\ref{log}). The latter alternative is the one used by most non-artificial stock traders today, and so subscription procedures and costs have already been carefully designed at most exchanges.

Agents place orders using the \texttt{Order} object, and its interface \texttt{OrderListener}~\cite{Sa02}. An order is created by calling the method \texttt{createOrder}. The \texttt{OrderListener} then implements the methods \texttt{orderCancelled} and \texttt{orderClosed}, which result in account updates and logging, as demonstrated in Fig.~\ref{log}. The account and portfolio maintenance should in any commercial implementation of the ATS be handled by standard F/X middleware. Similarly, the \texttt{AgentBroker} has only minimal functionality in this version of the ATS.

\section{Conclusions and Further Research}
\label{sec:conclusion}
We have described a proof-of-concept implementation of a running agent trade server, which we believe serves as a blueprint for future implementations intended for live use on a real financial exchange. We have also constructed and described an agent shell, currently in use for server implementation of agents of increasing complexity. These agent developers are also providing the first anecdotal evidence of the quality of our implementation. We intend to engage more developers before a formal evaluation is made. Our future work (cf. Fig.~\ref{gantt}) includes demonstrating that the size of an agent with a sophisticated business logic is manageable. We have also implemented various demonstrators for notifying services, to be used for presenting agent logs to the agent brokers and agent owners~\cite{NyByBo03}. While professional trading is a highly collaborative activity~\cite{HeJiLuHi93}, the agents in our set-up model each other only in a weak sense, based solely on their bidding. There are several multi-agent system aspects that deserve attention, such as multiple agent submission for teamwork, or even for manipulative purposes. Finally, the effects on the entire financial system as a result of wide adoption of agent trade is under investigation. This work is being done together with the Stockholm School of Economics and OM.

\section*{Acknowledgments}
The authors would like to thank Jesper Johansson and Michael Poijes, who wrote the first agent shell and some of the server code, and kindly provided log data. David Lyb\"ack and Ulf Essler contributed with important comments and fruitful discussions. The Vinnova project TAP on accessible autonomous software provided the authors with the time required for this study.




\end{document}